\documentclass[10pt,conference]{IEEEtran}
\IEEEoverridecommandlockouts
\usepackage{graphicx}
\usepackage[table]{xcolor}

\usepackage{cite}
\usepackage{glossaries}
\usepackage[hidelinks, bookmarks=false]{hyperref}
\usepackage[noabbrev,capitalise]{cleveref}
\usepackage{booktabs}
\usepackage{array}
\usepackage[symbol]{footmisc}
\usepackage{threeparttable}
\usepackage{makecell}
\usepackage{arydshln}
\usepackage{tikz}
\usepackage{eso-pic}
\usepackage{hyphenat}
\usepackage{listings}
\usepackage{microtype}
\usepackage{amsmath,amsfonts}
\usepackage{soul}
\usepackage{algorithm}
\usepackage{algpseudocode}
\usepackage{enumitem}
\usepackage{multirow}
\usepackage{xspace}
\usepackage{pifont}
\usepackage{changepage}
\usepackage{multicol}
\usepackage{amssymb}
\usepackage{printlen}
\usepackage{newtxtext,newtxmath}

\makeatletter
\newcommand\fs@spaceruled{%
  \def\@fs@cfont{\bfseries}%
  \let\@fs@capt\floatc@ruled%
  \def\@fs@pre{\vspace{.7\baselineskip}\hrule height.8pt depth0pt \kern2pt}%
  \def\@fs@post{\kern2pt\hrule\relax}%
  \def\@fs@mid{\kern2pt\hrule\kern2pt}%
  \let\@fs@iftopcapt\iftrue}
\makeatother

\usepackage[per-mode=repeated-symbol, free-standing-units, allow-number-unit-breaks]{siunitx}
\DeclareSIUnit\comp{COMP}
\DeclareSIUnit\flop{FLOP}
\DeclareSIUnit\flops{FLOPS}
\DeclareSIUnit\bps{bps}
\DeclareSIUnit\Bps{Bps}
\DeclareSIUnit\gate{GE}
\DeclareSIUnit\op{OP}
\DeclareSIUnit\macu{MACU}
\DeclareSIUnit\ops{OPS}
\DeclareSIUnit\core{core}
\DeclareSIUnit\request{request}
\DeclareSIUnit\cycle{cycle}
\DeclareSIUnit\teraops{TOPS}
\DeclareSIUnit\ghz{GHz}
\DeclareSIUnit\mhz{MHz}
\DeclareSIUnit[number-unit-product = ]\percent{\%}

\definecolor{MidnightBlue}{HTML}{191970}
\definecolor{Mint}{HTML}{3EB889}
\definecolor{EnglishRed}{HTML}{A4515C}
\definecolor{SelectiveYellow}{HTML}{FFBA08}
\definecolor{CyanProcess}{HTML}{08B2E3}
\definecolor{OliveDrab7}{HTML}{4D4730}
\definecolor{Red}{HTML}{FF0000}
\colorlet{color1}{MidnightBlue}
\colorlet{color2}{Mint}
\colorlet{color3}{EnglishRed}
\colorlet{color4}{SelectiveYellow}
\colorlet{color5}{CyanProcess}
\colorlet{color6}{OliveDrab7}
\colorlet{colorAlert}{Red}
\definecolor{PulpGreen}{HTML}{168638}
\definecolor{PulpBlue}{HTML}{1269b0}
\definecolor{PulpRed}{HTML}{a8322c}
\definecolor{PulpYellow}{HTML}{f2c100}

\Crefname{equation}{Eq.}{Eqs.}
\Crefname{figure}{Fig.}{Figs.}
\Crefname{tabular}{Tab.}{Tabs.}

\newacronym{ieee}{IEEE}{IEEE}
\glsunset{ieee}

\makeatletter \newcommand{\AddSpaceIfAnonymous}{\@ifclasswith{acmart}{anonymous}{\vspace{10mm}}{}} \makeatother
\PackageWarning{TODO:}{#1!}

\def\BibTeX{{\rm B\kern-.05em{\sc i\kern-.025em b}\kern-.08em
    T\kern-.1667em\lower.7ex\hbox{E}\kern-.125emX}}

\newacronym{ai}{AI}{Artificial Intelligence}
\newacronym{ml}{ML}{Machine Learning}
\newacronym{nn}{NN}{Neural Network}
\newacronym{cpu}{CPU}{Central Processing Unit}
\newacronym{asic}{ASIC}{Application Specific Integrated Circuit}
\newacronym[longplural={Systems-on-Chip}]{soc}{SoC}{System-on-Chip}
\newacronym{fpga}{FPGA}{Field Programmable Gate Array}
\newacronym{asip}{ASIP}{Application Specific Instruction Processor}
\newacronym{gpp}{GPP}{General Purpose Processor}
\newacronym{gp}{GP}{general-purpose}
\newacronym{gpgpu}{GP-GPU}{General Purpose Graphics Processing Unit}
\newacronym{gpu}{GPU}{Graphics Processing Unit}
\newacronym{sm}{SM}{Streaming Multiprocessor}
\newacronym{cuda}{CUDA}{Compute Unified Device Architecture}
\newacronym{mpi}{MPI}{Message Passing Interface}
\newacronym{cots}{COTS}{Commercial-Off-The-Shelf}
\newacronym{soa}{SoA}{state-of-the-art}
\newacronym{roi}{ROI}{Return on Investments}
\newacronym
[
  longplural={Core Complexes}
]
{cc}{CC}{Core Complex}

\newacronym{lte}{LTE}{Long Term Evolution}
\newacronym{nr}{NR}{New Radio}
\newacronym{4g}{4G}{4th Generation}
\newacronym{5g}{5G}{5th Generation}
\newacronym{b5g}{B5G}{Beyond-5G}
\newacronym{6g}{6G}{6th Generation}
\newacronym{urll}{URLL}{Ultra-Reliable Low-Latency}
\newacronym{mmtc}{mMTC}{massive Machine Type Communications}
\newacronym{embb}{eMBB}{enhanced Mobile Broadband}
\newacronym{3gpp}{3GPP}{3rd Generation Partnership Project}
\newacronym{oran}{O-RAN}{Open-RAN}
\newacronym{ran}{RAN}{Radio Access Networks}
\newacronym{cran}{C-RAN}{Cloud Radio Access Networks}
\newacronym{gnb}{gNB}{Next Generation Node B}
\newacronym{pusch}{PUSCH}{Physical Uplink Shared Channel}
\newacronym{sdr}{SDR}{Software Defined Radio}
\newacronym{phy}{PHY}{Physical}
\newacronym{cu}{CU}{Centralized Unit}
\newacronym{du}{DU}{Distributed Unit}
\newacronym{ru}{RU}{Remote Unit}
\newacronym{ue}{UE}{User Equipment}
\newacronym{ofdm}{OFDM}{Orthogonal Frequency Division Multiplexing}
\newacronym{ofdma}{OFDMA}{Orthogonal Frequency Division Multiple Access}
\newacronym{bf}{BF}{Beam Forming}
\newacronym{mimo}{MIMO}{Multiple-Input, Multiple-Output}
\newacronym{che}{CHE}{Channel Estimation}
\newacronym{dmrs}{DMRS}{Demodulation Reference Signal}
\newacronym{tti}{TTI}{Transmission Time Interval}
\newacronym{sc}{SC}{sub-carrier}
\newacronym{mu}{MU}{Multiple-User}
\newacronym{snr}{SNR}{Signal-to-Noise Ratio}
\newacronym{ber}{BER}{Bit-Error-Rate}

\newacronym{add}{add}{Add}
\newacronym{mul}{mul}{Multiply}
\newacronym{mac}{MAC}{Multiply\&Accumulate}
\newacronym{pmac}{p.mac}{Post-increment Multiply-accumulate}
\newacronym{axpy}{AXPY}{A Times X Plus Y}
\newacronym{dotp}{DOTP}{Dot Product}
\newacronym{sdotp}{SDOTP}{Sum Dot Product}
\newacronym{matmul}{MatMul}{Matrix Multiplication}
\newacronym{gemm}{GEMM}{General Matrix Multiplication}
\newacronym{mvm}{MVM}{Matrix-Vector Multiplication}
\newacronym{fft}{FFT}{Fast Fourier Transform}
\newacronym{sysinv}{SysInv}{Linear System Inversion}
\newacronym{choldec}{CholDec}{Cholesky Decomposition}
\newacronym{mmse}{MMSE}{Minimum Mean Squared Error}
\newacronym{conv2D}{Conv2D}{2D-Convolution}
\newacronym{dct}{DCT}{Direct Cosine Transform}

\newacronym{sram}{SRAM}{Static Random-Access Memory}
\newacronym{dram}{DRAM}{Dynamic Random-Access Memory}
\newacronym{spm}{SPM}{Scratchpad Memory}
\newacronym{tcdm}{TCDM}{Tightly Coupled Data Memory}
\newacronym{IDol}{I\$}{Instruction Cache}
\newacronym{dma}{DMA}{Direct Memory Access}
\newacronym{axi}{AXI}{Advanced eXtensible Interface}
\newacronym{noc}{NoC}{Nework on Chip}
\newacronym{csr}{CSR}{Control Status Register}
\newacronym{hbm}{HBM2E}{High Bandwidth Memory}

\newacronym{ipc}{IPC}{instructions-per-cycle}
\newacronym{wfi}{WFI}{wait-for-interrupt}
\newacronym{raw}{RAW}{read-after-write}
\newacronym{ins}{INS}{instruction}

\newacronym{fpu}{FPU}{Floating Point Unit}
\newacronym{fpss}{FP-SS}{Floating Point Sub-System}
\newacronym{ipu}{IPU}{Integer Processing Unit}
\newacronym{divsqrt}{DIVSQRT}{Division and Square-Root Unit}
\newacronym{lsu}{LSU}{Load Store Unit}
\newacronym{dsp}{DSP}{Digital Signal Processing}
\newacronym{qlr}{QLR}{Queue-Linked Register}

\newacronym{eda}{EDA}{Electronic Design Automation}
\newacronym{ge}{GE}{Gate Equivalent}
\newacronym{fo4}{FO4}{Fan-Out-of-4}
\newacronym{beol}{BEOL}{Back-End-of-Line}
\newacronym{pnr}{PnR}{Place and Route}
\newacronym{ppa}{PPA}{Power, Performance and Area}

\newacronym{numa}{NUMA}{Non-Uniform Memory Access}
\newacronym{fc}{FC}{Fully-Connected}
\newacronym{isa}{ISA}{Instruction Set Architecture}
\newacronym{simd}{SIMD}{Single Instruction Multiple Data}
\newacronym{spmd}{SPMD}{Single Program Multiple Data}
\newacronym{cdf}{CDF}{Cumulative Distribution Function}
\newacronym{api}{API}{Application Programmable Interface}
\newacronym{rtl}{RTL}{Register Transfer Level}
\newacronym{sfr}{SFR}{Synchronization Free Region}
\newacronym{dsl}{DSL}{Domain-Specific Language}

\newacronym{int}{INT}{integer}
\newacronym{fp}{FP}{floating-point}

\newacronym{pe}{PE}{Positional Encoding}
\newacronym{rg}{RG}{Resource Grid}
\newacronym{re}{RE}{Resource Element}
\newacronym{cp}{CP}{Cyclic Prefix}
\newacronym{llr}{LLR}{Log-Likelihood Ratio}
\newacronym{lmmse}{LMMSE}{Linear Minimum Mean Squared Error}
\newacronym{3d}{3D}{3-Dimensional}
\newacronym{2d}{2D}{2-Dimensional}
\newacronym{prb}{PRB}{Physical Resource Block}
\newacronym{mdx}{MDX}{Model Driven Neural Receiver}
\newacronym{dals}{DA-LS}{Data Aided Least Squares}
\newacronym{pals}{PA-LS}{Pilot Aided Least Squares}
\newacronym{cdm}{CDM}{Code-Division Multiplexing}
\newacronym{ldpc}{LDPC}{Low Density Parity Check Code}
\newacronym{bce}{BCE}{Binary Cross-Entropy}
\newacronym{mse}{MSE}{Mean Squared Error}
\newacronym{adam}{ADAM}{Adaptive Moment Estimation}
\newacronym{umi}{UMi}{Urban Microcell}
\newacronym{mcs}{MCS}{Modulation Coding Scheme}
\newacronym{tdl}{TDL}{Tapped Delay Line}
\newacronym{ls}{LS}{Least Squares}
\newacronym{tbler}{TBLER}{Transport Block Error Rate}
\newacronym{cscs}{CSCS}{Swiss National Supercomputing Centre}
\newacronym{flop}{FLOP}{Floating Point Operation}
\newacronym{resblock}{ResBlock}{ResNet block}
\newacronym{qam}{QAM}{Quadrature Amplitude Modulation}

\author{%
Mahdi Abdollahpour\textsuperscript{\ensuremath{\S}}\quad
Marco Bertuletti\textsuperscript{\ensuremath{*}}\quad
Yichao Zhang\textsuperscript{\ensuremath{*}}\quad
Yawei Li\textsuperscript{\ensuremath{*}}\quad \\
Luca Benini\textsuperscript{\ensuremath{\S*}}\quad
Alessandro Vanelli-Coralli\textsuperscript{\ensuremath{\S*}}
\\
{\small
 \textsuperscript{\ensuremath{\S}}DEI, University of Bologna\quad
 \textsuperscript{\ensuremath{*}}IIS, ETH Z\"{u}rich
}
\\
{\small\itshape%
\textsuperscript{\ensuremath{\S}}\{mahdi.abdollahpour,luca.benini,alessandro.vanelli\}@unibo.it, \textsuperscript{\ensuremath{*}}\{mbertuletti,yiczhang,yawli,lbenini,avanelli\}@iis.ee.ethz.ch %
}
}

\floatstyle{spaceruled}
\restylefloat{algorithm}

\begin{document}
\title{A Compute\&Memory Efficient Model-Driven Neural 5G Receiver for Edge AI-assisted RAN\\
}
\maketitle


\begin{abstract}
Artificial intelligence approaches for base-band processing for radio receivers have demonstrated significant performance gains. 
Most of the proposed methods are characterized by high compute and memory requirements, hindering their deployment at the edge of the Radio Access Networks (RAN) and limiting their scalability to large bandwidths and many antenna 6G systems.
In this paper, we propose a low-complexity, model-driven neural network-based receiver,  designed for multi-user multiple-input multiple-output (MU-MIMO) systems and suitable for implementation at the RAN edge. 
The proposed solution is compliant with the 5G New Radio (5G NR), and supports different modulation schemes, bandwidths, number of users, and number of base-station antennas with a single trained model without the need for further training.
Numerical simulations of the Physical Uplink Shared Channel (PUSCH) processing show that the proposed solution outperforms the state-of-the-art methods in terms of achievable  Transport Block Error Rate (TBLER), while reducing the Floating Point Operations (FLOPs) by 66$\times$, and the learnable parameters by 396$\times$.
\end{abstract}


\section{Introduction}
The evolution of \gls{b5g} and 6G \gls{ran}, enabled by diverse edge functions, is rapidly advancing the network's service quality, capabilities, and user densities in complex deployment scenarios~\cite{saad2019vision}.
With the evolution of functional disaggregation in \gls{3gpp} and \gls{oran}, \gls{ai}-based \gls{ran} processing is emerging as a key trend, promoted by open interfaces~\cite{lin2023artificial}.
Processing the uplink \gls{phy}-layer at the \gls{ran} edge is crucial for improving latency, performance, and system flexibility, but it stands out as one of the most computational and memory demanding \gls{ran} functions~\cite{bjornson2017massive}.

Recent research~\cite{zhang2020artificial} has focused on \gls{ai}-for-\gls{ran} to enhance \gls{phy}-layer performance. \gls{nn}-based \gls{ofdma} receivers have demonstrated improved \gls{ber} performance compared to conventional \gls{lmmse}-receivers~\cite{honkala2021deeprx,korpi2021deeprx,wiesmayr2024design,cammerer2023neural}.
However, these models incur high computational complexity and large memory footprint compared to classical approaches, exacerbating the computational bottleneck of the \gls{phy} processing.

As a consequence of growing computational requirements for mixed \gls{ai}\&wireless workloads, base-station edge-processors are evolving from \gls{ran}-specialized \glspl{asic} to high-performance many-core programmable processors\cite{Marvell_Octeon10,Zhang_TeraPool_2024,Quallcom_X100, NvidiaH100ArchWP, IntelArchDay2021SPR}.

The top three rows of table~\ref{tab:archs} provide an insight into the computing capability, on-chip memory of these devices relative to the performance and memory footprint required by a runtime-compliant \gls{soa} neural receiver (NRX)~\cite{wiesmayr2024design}, addressing a 4$\times$2 \gls{mu}-\gls{mimo} problem (4 receive antennas, 2 transmit data streams) in a 1-ms \gls{tti} allocating 273 \glspl{prb} per stream. 
None of the devices intended for edge deployment can provide the required performance (as indicated by the required-to-processor performance ratios). 
High performance CPUs and GPUs have significantly higher capabilities, but they vastly exceed the power budget of a base station edge-processor. Hence, they need to be accessed via the fronthaul link of the \gls{gnb}, causing a significant increase of end-to-end latency. 
\begin{table}[h!]
    \vspace{-6pt} 
    \caption{Increase in computational complexity and memory capacity required to SoA RAN processors for Edge deployment of NRX}
    \vspace{-8pt} 
    \label{tab:archs}
    \centering
    \setlength{\tabcolsep}{1pt} 
    \renewcommand{\arraystretch}{1} 
    \resizebox{\columnwidth}{!}{
    \begin{tabular}{l c@{\hspace{5pt}} p{1cm} p{1cm}@{\hspace{-5pt}} p{1cm}@{\hspace{-4pt}} p{1cm}@{\hspace{-2pt}} p{1cm} p{1cm}} 
        \hline
         & Edge & 16b-TFLOPs & RAM (MiB) & Power (W) & Req.-Perf. & Req.-Mem.$^{\mathrm{a}}$ \\
        \hline
        \hline
        Marvell OCTEON10~\cite{Marvell_Octeon10}  & yes & 1     & 24 & 50    & 75.86$\times$ & 1.09$\times$ \\
        TeraPool~\cite{Zhang_TeraPool_2024}       & yes & 3.7   & 4  & 7     & 20.86$\times$ & 6.54$\times$ \\
        Qualcomm X100~\cite{Quallcom_X100}       & yes & -     & -     & 18    & -           & - \\
        \hline
        NVDIA H100~\cite{NvidiaH100ArchWP}              & no & 1979  & 50    & 510 & 0.04$\times$ & 0.52$\times$ \\
        Intel Sapphire Rapids~\cite{IntelArchDay2021SPR}   & no & 75.3 & 112.5 & -   & 1.04$\times$ & 0.23$\times$ \\
        \hline
        \multicolumn{6}{l}{$^{\mathrm{a}}$Size of model parameters plus inputs.}
    \end{tabular}
    }
\end{table}
\vspace{-3pt} 

This status quo stresses the need for lightweight models tailored for edge deployment that curtail compute and memory resources, preserving critical \gls{ran} performance metrics.

The \gls{soa} neural receivers use a fully data-driven approach which leads to large models with high computational costs~\cite{cammerer2023neural,wiesmayr2024design,honkala2021deeprx}. In contrast, in our solution we follow a model-driven approach: we augment the conventional \gls{lmmse} receiver with learnable parameters, and specialized \glspl{resblock} processing to suppress noise and interference, which results in significant complexity reduction.

In this paper, we propose the design of a small, efficient \gls{mdx}, fully compatible with the \gls{5g} \gls{nr} and suitable for edge-\gls{ran} deployment. 
The proposed design is available open-source~\cite{open_source}.
The main contributions of the proposed design are:
\begin{itemize}
    \item The combined design of conventional baseband processing blocks with learnable parameters and specialized \glspl{resblock} processing to suppress noise and interference, approaching \gls{tbler} performance near the limit with perfect channel knowledge.
    \item The evaluation of the receiver performance for the \gls{pusch} in terms of \gls{tbler} in \gls{3gpp} \gls{tdl}-A channel model.
    \item The assessment of the computational complexity, and the memory footprint of our model against \gls{soa} AI-receivers.
\end{itemize}
Our proposed \gls{mdx} delivers superior \gls{tbler} performance and also achieves a 66$\times$ reduction in \glspl{flop} and a 396$\times$ reduction in model parameters compared to the purely data-driven \gls{soa}, making it well suited for edge deployment.
Furthermore, despite being trained only on a small 4$\times$2 MIMO problem, we show that \gls{mdx} effectively generalizes to support larger MIMO sizes and diverse configurations (e.g., modulation schemes, bandwidths), facilitating practical online training on edge devices.

\vspace{-3pt} 

\section{System Model}
\begin{figure}[tbp] 
\centerline{\includegraphics[width=.9\columnwidth]{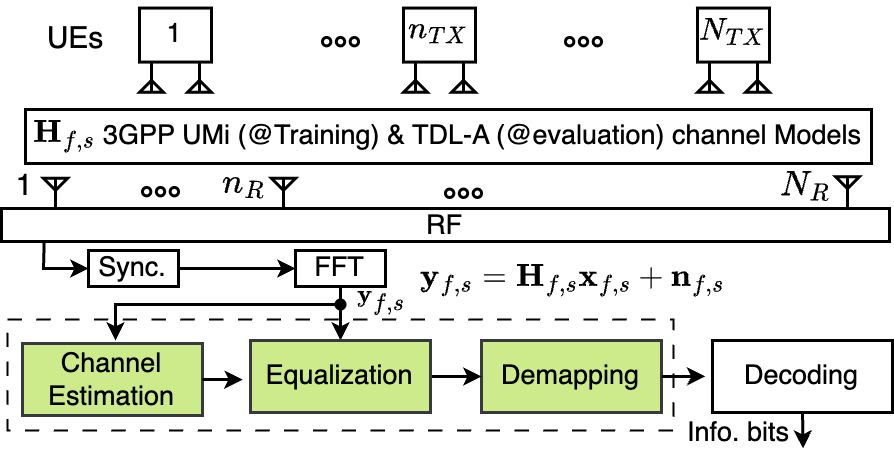}}
\vspace{-15pt} 
\caption{The Mu-MIMO uplink system model. The colored conventional processing blocks are replaced with the proposed \gls{mdx} model.}
\label{fig:phy}
\end{figure}

\vspace{-4pt} 
This section describes the telecommunication system (Fig.~\ref{fig:phy}) used for our model evaluation.
We consider the \gls{5g} \gls{nr} \gls{mu}-\gls{mimo} \gls{ofdm} uplink transmission, where $N_{TX}$ layers (data streams) from $N_U$ \glspl{ue} are transmitted to a \gls{gnb} with $N_R$ receiving antennas. 
Each \gls{ue} may have multiple antennas and multiple layers: 
we generalize to multiple antennas by assuming 2 transmitting antennas and 1 layer per \gls{ue}.

Each layer's transmission is divided into time slots. Within each slot the data is mapped on a \gls{rg}: the set $\mathcal{RG} \! = \! \{1, \dots, F\}\! \times\! \{1, \dots, S\}$, where $F$ is the number of subcarriers, and $S$ is the number of OFDM symbols in a slot. The smallest RG unit is a \gls{re}, identified by a subcarrier index $f$ and an OFDM symbol index $s$, $(f,s) \in \mathcal{RG}$. Each \gls{re}, associated with the layer $n_{TX}$, where $1\!\leq\! n_{TX}\! \leq\! N_{TX}$, carries a complex symbol $x_{f,s,n_{TX}}$. The symbol $x_{f,s,n_{TX}}$ is defined by a $2^{B_{n_{TX}}}$ \gls{qam} constellation, and encodes a vector of bits $\mathbf{b}_{f,s,n_{TX}}\!\in\!\{0,1\}^{B_{n_{TX}}}$, where ${B_{n_{TX}}}$ denotes \gls{qam} order of layer $n_{TX}$. Slots typically contain 14 OFDM symbols, and the frequency domain is organized into \glspl{prb}, each \gls{prb} comprising 12 subcarriers, which serve as the fundamental units for resource allocation~\cite{3gpp_ts_38211}.

After the FFT, the received signal on RE $(f,s)$ is:
\vspace{-4pt} 
\begin{equation}
    \mathbf{y}_{f,s}=\mathbf{H}_{f,s}\mathbf{x}_{f,s}+\mathbf{n}_{f,s},
    \label{eq:system_model}
\vspace{-4pt} 
\end{equation}
where $\mathbf{y}_{f,s} \! \in \! \mathcal{C}^{N_{R}}$, and $\mathbf{x}_{f,s} \! \in \! \mathcal{C}^{N_{TX}}$ are the received and transmitted symbols, $\mathbf{H}_{f,s}\! \in \! \mathcal{C}^{N_R \times N_{TX}}$ is the \gls{mimo} channel matrix, and $\mathbf{n}_{f,s}\! \in \! \mathcal{C}^{N_R}$ is the complex additive Gaussian noise with power spectral density $N_0$, distributed as $\mathcal{CN}(\mathbf{0},N_0\mathbf{I})$. 

The majority of \glspl{re} carry data symbols. A subset within each \gls{rg} transmits the \glspl{dmrs}, also known as pilots. The data-carrying positions can be defined as the set \( \mathcal{D} \), including all \gls{re} indices \((f, s)\) corresponding to data symbols. \glspl{dmrs} and their positions within an \gls{rg} are layer-specific, known at the receiver, and used for channel estimation. The \gls{dmrs} positions for layer \( n_{TX} \) are defined by the set \( \mathcal{P}_{n_{TX}} \) of all the \((f, s)\) indices of \glspl{re} carrying pilots. In our configuration, the \gls{dmrs} positions are located at \gls{ofdm} symbols 2 and 11 and are configured to use a \gls{cdm} group size of 2 in the frequency domain.

\vspace{-2pt} 

\section{Model Driven Neural Receiver (MDX)}
\vspace{-1pt} 

\begin{figure}[tbp] 
\centerline{\includegraphics[width=0.9\columnwidth]{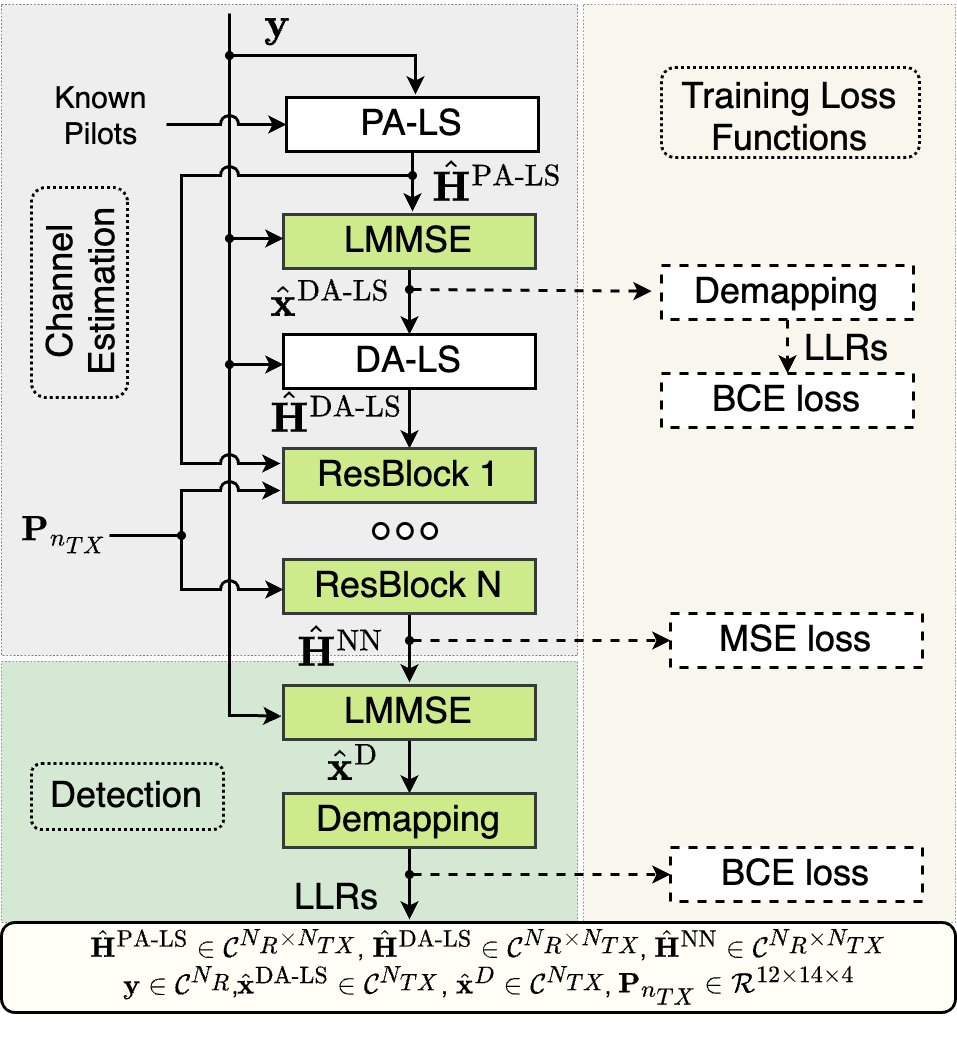}}
\vspace{-14pt} 
\caption{Model driven neural receiver block diagram. The colored blocks include trainable weights. The dashed blocks are used only in training.}
\label{fig:framework}
\end{figure}

In this section, we describe the architecture of our lightweight model: a mix of \gls{ai}-enhanced classical signal processing blocks and \gls{ai} blocks. 
The overall architecture of \gls{mdx} is shown in Fig.~\ref{fig:framework}. 
The model applies first \gls{pals} channel estimation, then \gls{lmmse} equalization with learnable parameters. The core of the \gls{mdx} network consists of \gls{dals} channel estimation, specialized \glspl{resblock}, and a Demapper with learnable parameters, which outputs the final bit \glspl{llr}. 
Finally, the \glspl{llr} undergo \gls{ldpc} decoding. We detail these steps in the following.

\subsection{\gls{pals} Channel Estimation}

In \gls{5g} \gls{mu}-\gls{mimo}, pilot sequences assigned to different layers are orthogonal in time, frequency, or code domains. These predefined signals are used to obtain an initial estimate of channel vectors at the \gls{re} positions carrying pilots as
\begin{equation}    
\hat{\mathbf{h}}_{f,s,n_{TX}}^\text{PA-LS}=\frac{p^*_{f,s,n_{TX}} \mathbf{y}_{f,s}}{|p_{f,s,n_{TX}}|^2}\in \mathcal{C}^{N_R}, \quad \forall [f,s] \in \mathcal{P}_{n_{TX}},
\label{eq:pa_ls}
\end{equation}
where \( p_{f,s,n_{TX}}\) denotes the complex-valued pilot symbol transmitted by layer \( n_{TX} \) at \gls{re} \((f,s)\), \( \mathbf{y}_{f,s} \) represents the corresponding received signal, and the notation \( (\cdot)^* \) denotes the complex conjugate. When \gls{cdm} groups are used, the pilots of layers in the same \gls{cdm} group are not orthogonal in time and in frequency. In such cases, the \gls{pals} channel estimate in \eqref{eq:pa_ls} is averaged over the subcarriers in the same group. Then, linear interpolation is used to obtain the channel estimates at data-carrying \glspl{re}. The \gls{pals} channel vectors can be arranged in matrix form as 
\begin{equation}
\hat{\mathbf{H}}_{f,s}^{\text{PA-LS}} = 
\left[ \hat{\mathbf{h}}_{f,s,1}^{\text{PA-LS}}, \dots, \hat{\mathbf{h}}_{f,s,N_{TX}}^{\text{PA-LS}} \right] \in \mathcal{C}^{N_R \times N_{TX}}.
\label{eq:pa_ls_mat}
\end{equation}

\subsection{Equalization and Demapping}

We use the \gls{lmmse} equalization to obtain an estimate of the transmitted symbols. The \gls{lmmse} matrix is defined as
\vspace{-3pt} 
\begin{equation}    
\mathbf{G}_{f,s}=\left( \hat{\mathbf{H}}_{f,s}^H \hat{\mathbf{H}}_{f,s}+\hat{\sigma}^2_{\text{adj},f,s}I \right)^{-1} \hat{\mathbf{H}}_{f,s}^H \in \mathbf{C}^{N_{TX} \times N_R},  
\label{eq:lmmse_mat}
\end{equation}
where $\hat{\mathbf{H}}_{f,s} \in \mathcal{C}^{N_R \times N_{TX}}$ is an estimate of the MIMO channel. The superscript $H$ denotes Hermitian operation, $I$ is a $N_{TX} \times N_{TX}$ identity matrix, and
\vspace{-1pt} 
\begin{equation}
\hat{\sigma}^2_{\text{adj},f,s} = \mathbf{\Psi}(f^{\prime}, s) \cdot \hat{\sigma}_{f,s},
\label{eq:lmmse_input_noise_mul}
\end{equation}
where $\hat{\sigma}_{f,s}$ is the estimated noise variance, $f^{\prime}\!=\!((f-1)\! \bmod 12)\! +\! 1$, '$\bmod$' being modulo operation, $\mathbf{\Psi}\! \in\! \mathcal{R}^{12 \times 14}$ is a learnable matrix. Due to the sparse placement of pilot symbols within an \gls{rg}, which provide only localized channel information, the channel estimation error varies across \glspl{re} near the pilot locations and those farther away. The learnable matrix $\mathbf{\Psi}$ adjusts the \gls{lmmse} equalizer’s input error variances on a per-\gls{prb} basis, enabling the receiver to adapt to these variations. The equalized symbols can be computed as
\begin{equation}    
\hat{\mathbf{x}}_{f,s}= \text{Diag}(\mathbf{G}_{f,s}\hat{\mathbf{H}}_{f,s})^{-1} \mathbf{G}_{f,s}\mathbf{y}_{f,s} \in \mathcal{C}^{N_{TX}}.  
\label{eq:lmmse_eqz}
\end{equation}
where the operator \(\text{Diag}(\cdot)\) constructs a diagonal matrix using the diagonal elements of the input matrix. The effective post-equalization residual noise variances are
\begin{equation}    
\hat{\boldsymbol{\sigma}}_{\text{res},f,s}= \text{diag} \left( \text{Diag}(\mathbf{G}_{f,s}\hat{\mathbf{H}}_{f,s})^{-1} -\mathbf{I} \right).  
\label{eq:lmmse_noise_out}
\end{equation}
where, the operator \(\text{diag}(\cdot)\) returns the diagonal elements of the input matrix as a vector, and $\hat{\boldsymbol{\sigma}}_{\text{res},f,s} \! = \! \{ \hat{\sigma}^2_{\text{res},f,s,1} , \dots, \hat{\sigma}^2_{\text{res},f,s,N_{TX} } \}$. The equalization process can be written as the function
\begin{equation}
    \hat{\mathbf{x}}_{f,s}, \hat{\boldsymbol{\sigma}}_{\text{res},f,s} =  \text{LMMSE}_{\mathbf{\Psi}}\big(\hat{\mathbf{H}}_{f,s}, \mathbf{y}_{f,s}, \hat{\sigma}_{f,s}^2 \big).
    \label{eq_fn_lmmse}
\end{equation}
Lets define \gls{llr} for a bit $q$ as $L \! = \! \log(p/(1-p))$, where $p$ denotes the probability of $q\!=\! 1$. Then the estimated symbols can be demapped onto \glspl{llr} as
\begin{equation}    
L_b=\frac{1}{\hat{\sigma}^2_{\text{dem}}} ( \arg \min_{x \in \mathcal{C}_b^0} ||\hat{x}-x||_2^2 - \arg \min_{x \in \mathcal{C}_b^1} ||\hat{x}-x||_2^2  ),
\label{eq:llrs}
\end{equation}
where the sets \(\mathcal{C}_b^0\) and \(\mathcal{C}_b^1\) represent the constellation points for which the \(b\)-th bit is 0 and 1, respectively, and $b=1,\dots,B$ with $B$ denoting the number of bits per symbol. The noise variance is adjusted as
\begin{equation}
\hat{\sigma}^2_{\text{dem},f,s} = \gamma_{m} \mathbf{\Phi}(f^{\prime}, s) \cdot \hat{\sigma}_{\text{res},f,s}^2,
\label{eq:demapper_input_noise_mul}
\end{equation}
where $\mathbf{\Phi}\in \mathcal{R}^{12 \times 14}$ is a learnable matrix, and $\gamma_{m}$ for $m=1,\dots,M$ is a modulation-specific learnable scaler where $M$ indicates the number of supported modulation orders. The demapping function can be written as 
\begin{equation}
    L_{b,f,s,n_{TX}} \! = \! \text{Demapper}_{\gamma_m,\mathbf{\Phi}}\! \left(\hat{\mathbf{x}}_{f,s}(n_{TX}), \hat{\boldsymbol{\sigma}}_{\text{res},f,s}(n_{TX})\right)
    \label{eq:fn_demapping}
\end{equation}

\subsection{\gls{dals} Channel Estimation}

For all $(f,s) \in \mathcal{D}$, the data symbols can be estimated using \gls{pals} channel matrix as
\begin{equation}
    \hat{\mathbf{x}}_{f,s}^\textit{DA-LS}, \hat{\boldsymbol{\sigma}}_{\textit{res},f,s}^\textit{DA-LS} =  \text{LMMSE}_{\mathbf{\Psi}^\textit{DA-LS}}\big(\hat{\mathbf{H}}_{f,s}^\textit{PA-LS}, \mathbf{y}_{f,s}, \hat{\sigma}_{f,s}^2\big).
    \label{eq:x_dals}
\end{equation}
Then the estimated symbols in \eqref{eq:x_dals} can be used to obtain a data-aided estimate of channel as
\begin{equation}
\hat{\mathbf{h}}_{f,s,n_{TX}}^{\text{DA-LS}} \! = \! 
\left( \mathbf{y}_{f,s} \! - \! \mathbf{\hat{H}}_{f,s,\backslash n_{TX}}^{\text{PA-LS}}  
\mathbf{\hat{x}}_{f,s,\backslash n_{TX}}^{\text{DA-LS}} \right) \! \cdot \! (\hat{\mathbf{x}}_{f,s}^{\text{DA-LS}}(n_{TX}))^*, 
\label{eq:da_ls}
\end{equation}
where, $\hat{\mathbf{h}}_{f,s,n_{TX}}^{\text{DA-LS}} \in \mathcal{C}^{N_R}$, the subscript $(\cdot)_{\backslash n_{TX}}$ excludes the $n_{TX}$-th column or element from a matrix or a vector respectively. 
The resulting \gls{dals} channel vectors in matrix form are
\begin{equation}
\hat{\mathbf{H}}_{f,s}^{\text{DA-LS}} = 
\left[ \hat{\mathbf{h}}_{f,s,1}^{\text{DA-LS}}, \dots, \hat{\mathbf{h}}_{f,s,N_{TX}}^{\text{DA-LS}} \right] \in \mathcal{C}^{N_R \times N_{TX}}.
\label{eq:da_ls_mat}
\end{equation}

Note that, the resulting channel estimate in \eqref{eq:da_ls} has different scaling from the actual channel. Additionally, it is noisy, and retains post-equalization residual interferences. However, the following neural processing, using our specialized ResNet blocks, can learn to adjust the scale and suppress noise and interference.

\subsection{\glspl{resblock} Processing}

\begin{figure}[tbp] 
\centerline{\includegraphics[width=0.95\columnwidth]{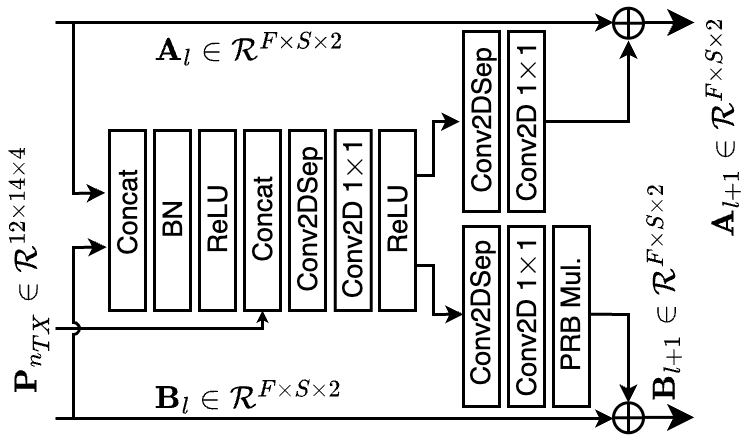}}
\vspace{-14pt} 
\caption{The specialized \gls{resblock}. The "Conv2DSep" and "Conv2D 1\(\times\)1" form a depthwise separable convolution~\cite{chollet2017xception}, and "BN" indicates batch normalization.}
\label{fig:chnn_resblock}
\end{figure}

The so far estimated \gls{mimo} channel is enhanced by further processing through a series of N \glspl{resblock}. 
The block diagram of our specialized \gls{resblock} is illustrated in Fig.~\ref{fig:chnn_resblock}. 
In the figure, $A_l$ and $B_l$ are the outputs of previous \gls{resblock}, and $P_{n_{TX}}$ is the positional encoding of layer $n_{TX}$. 
The block "PRB Mul." indicates a per-PRB weighting multiplier. The details of \glspl{resblock} processing is explained in the following.

Defining the set of \gls{mimo} links as \( \mathcal{L}\! =\! \{\!1,\!\dots\!,\!N_{\!R}\!\}\! \times \! \{ \! 1,\!\dots\!,\!N_{\!TX}\!\} \), the \glspl{resblock} process every \(\! (n_R,n_{TX})\! \in \! \mathcal{L} \), in parallel using shared weights. The inputs of the network are:

1. The \gls{pals} channel estimates in \eqref{eq:pa_ls_mat} reshaped as a \gls{3d} tensor $\mathbf{A}_1 \! \in \! \mathcal{R}^{\!F\! \times S \times 2}$, with its $(f,s,k)$-th element defined as
\begin{equation}
\mathbf{A}_1(f, s, k) = 
\begin{cases} 
\Re\left( \hat{H}_{f,s}^{\text{PA-LS}} (n_R,n_{TX}) \right) & \text{if } k = 1, \\
\Im\left( \hat{H}_{f,s}^{\text{PA-LS}} (n_R,n_{TX}) \right) & \text{if } k = 2,
\end{cases}
\label{eq:resblock_a}
\end{equation}
\vspace{-7pt} 

2. The \gls{dals} channel estimates in \eqref{eq:da_ls_mat} reshaped as a \gls{3d} tensor $\mathbf{B}_1 \in \mathcal{R}^{F \times S \times 2}$ with its $(f,s,k)$-th element defined as
\vspace{-2pt} 
    \begin{equation}
    \mathbf{B}_1(f, s, k) = 
    \begin{cases} 
    \Re\left( \hat{H}_{f,s}^{\text{DA-LS}} (n_R,n_{TX}) \right) & \text{if } k = 1, \\
    \Im\left( \hat{H}_{f,s}^{\text{DA-LS}} (n_R,n_{TX}) \right) & \text{if } k = 2,
    \end{cases}
    \label{eq:resblock_b}
    \end{equation}
\vspace{-5pt} 

3. The layer-specific \gls{pe}, $\mathbf{P}_{\!n_{TX}}\! \in \! \mathcal{R}^{\!12 \times 14 \times 4}$, which is provided to all \glspl{resblock}. We incorporate a per-\gls{prb} positional encoding to capture the channel's structure across time and frequency domains. The encoding consists of two components. The first component $\dot{\mathbf{P}}$, inspired by~\cite{cammerer2023neural}, consists of the normalized relative vertical (in frequency) and horizontal (in time) distance of each \gls{re} inside a \gls{prb} from the nearest \gls{dmrs}. It is a real-valued \gls{3d} tensor with a shape of $12\! \times \! 14 \! \times \! 2$. Since the \gls{dmrs} positions are layer-specific, this component of the \gls{pe} is also layer-dependent. The second component, $\ddot{\mathbf{P}}$, is a real-valued \gls{3d} tensor of shape $12\! \times\! 14 \! \times \! 2$, consisting of the normalized absolute cartesian coordinates of each \gls{re} inside a \gls{prb}, and is defined as
\vspace{-3pt} 
    \begin{equation}
    \! \ddot{\mathbf{P}}({f_p,s_p,k})\!=\! \begin{cases} \frac{f_p}{12}\! &\! \text{if } k\! =\! 1, \\ \frac{s_p}{14}\! &\! \text{if } k\! =\! 2, \end{cases}\!,\! \quad \! \forall 1\!\leq \!f_p\! \leq \! 12,1\!\leq \! s_p\! \leq \! 14.\!
    \label{eq:pe_abs}
    \end{equation} 
    \vspace{1pt} 
    Then, the overall \gls{pe} is created by concatenating the two components, $\dot{\mathbf{P}}$ and $\ddot{\mathbf{P}}$, along the last dimension.

Algorithm~\ref{alg:resblock} summarizes the \glspl{resblock} processing. We use $N\!=\!4$ \glspl{resblock}, where each \gls{resblock} produces two outputs, A and B, except for the final block, which generates only A. The A outputs are weighted on a per-PRB basis before performing the residual summation, using a learnable multiplier \(\mathbf{\Gamma}_l \in \mathbb{R}^{12 \times 14 \times 1}\), $l=1,\dots,N$. The tensors are repeated and broadcast to the appropriate shapes when necessary. The network output for the link \((n_R, n_{TX})\) is denoted by \(\mathbf{A}_{N+1} \in \mathbb{R}^{F \times S \times 2}\). This output can be interpreted as a complex-valued matrix \(\mathbf{C}_{n_R, n_{TX}} \in \mathbb{C}^{F \times S}\).


\begin{algorithm}[tbp]

\small

\caption{\glspl{resblock} Processing}
\label{alg:resblock}

\begin{algorithmic}[1]

\State \textbf{Inputs:} $\hat{H}_{f,s}^{\text{PA-LS}}\!, \hat{H}_{f,s}^{\text{DA-LS}}$ $\forall (\!f,s\!)\!\in\!\mathcal{RG}$, $\mathbf{P}_{n_{TX}}$ $\forall 1\!\le\! n_{TX}\!\le\! N_{TX}$, $N$.

\For{$(n_R,n_{TX}) \in \mathcal{L}$}
\State Create $\mathbf{A}_1$, $\mathbf{B}_1$ \Comment{\eqref{eq:resblock_a}, \eqref{eq:resblock_b}}
\State $\mathbf{P} \gets \text{Repeat } \mathbf{P}_{n_{TX}} \text{ on first dimension } $ \Comment{$F \times S \times 4$}
\For{$l \gets 1$ to $N$}
    \State $\mathbf{X} \gets \text{Concat}(\mathbf{A}_l, \mathbf{B}_l)$ \Comment{$F \times S \times 4$}
    \State $\mathbf{X} \gets \text{ReLU}\left( \text{BN}( \mathbf{X} )  \right)$ \Comment{$F \times S \times 4$}
    \State $\mathbf{X} \gets \text{Concat}(\mathbf{X}, \mathbf{P})$ \Comment{$F \times S \times 8$}
    \State $ \mathbf{X} \gets \text{Conv2D}\left(  \text{Conv2DSep}(\mathbf{X})\right)$ \Comment{$F \times S \times 8$}
    \State $\mathbf{X} \gets \text{ReLU}(\mathbf{X})$ \Comment{$F \times S \times 8$}
    \State   $ \mathbf{A}_{l+1} \gets \text{Conv2D}(\text{Conv2DSep}(\mathbf{X}))$ \Comment{$F \times S \times 2$}
    \State   $\tilde{\mathbf{\Gamma}}_l \gets \text{Repeat and Broadcast } \mathbf{\Gamma}_l$ \Comment{$F \times S \times 2$}
    \State   $\mathbf{A}_{l+1} \gets \mathbf{A}_{l} + \tilde{\mathbf{\Gamma}}_l\cdot \mathbf{A}_{l+1}$ \Comment{$F \times S \times 2$}

    \If{$l < N$}
    \State  $\mathbf{B}_{l+1} \gets \text{Conv2D}(\text{Conv2DSep}(\mathbf{B}_{l+1}))$ \Comment{$F \times S \times 2$}
    \State   $\mathbf{B}_{l+1} \gets \mathbf{B}_{l} + \mathbf{B}_{l+1}$ \Comment{$F \times S \times 2$}
    \EndIf
\EndFor
\State $\mathbf{C}_{n_R,n_{TX}} \gets \text{Complex}(\mathbf{A}_{N+1})$ \Comment{$\mathbf{C}_{n_R, n_{TX}} \in \mathbb{C}^{F \times S}$}
\EndFor
\State \textbf{Output:} $\mathbf{C}_{n_R,n_{TX}}, \quad \forall (n_R,n_{TX}) \in \mathcal{L}$ 
\end{algorithmic}
\end{algorithm}

\vspace{-4pt} 

\subsection{Detection}
\vspace{-3pt} 

We use the channel estimates enhanced by \glspl{resblock} processing for the MIMO detection. The output of Alg.~\ref{alg:resblock} can be written as $\hat{\mathbf{H}}^\text{NN}_{f,s} \in \mathcal{C}^{N_R \times N_{TX}}$ where its $(n_R,n_{TX})$-th element defined as
\vspace{-3pt} 
\begin{equation}
    \hat{\mathbf{H}}^\text{NN}_{f,s}(n_R,n_{TX})=\mathbf{C}_{n_R,n_{TX}}(f,s).
    \label{eq:h_nn}
    \vspace{-4pt} 
\end{equation}
The equalized symbols and the processed noise variance are
\vspace{-2pt} 
\begin{equation}
    \hat{\mathbf{x}}_{f,s}^\text{D}, \hat{\boldsymbol{\sigma}}_{\text{res},f,s}^\text{D} =  \text{LMMSE}_{\mathbf{\Psi}^\text{D}}\big(\hat{\mathbf{H}}_{f,s}^\text{NN}, \mathbf{y}_{f,s}, \hat{\sigma}_{f,s}\big).
    \label{eq:x_detection}
    \vspace{-2pt} 
\end{equation}
Then the \glspl{llr} can be computed as
\vspace{-4pt} 
\begin{equation}
    L_{b,f,s,n_{TX}}^{\text{D}} = \text{Demapper}_{\gamma_m,\mathbf{\Phi}} \left(\hat{\mathbf{x}}_{f,s}^\text{D}(n_{TX}), \hat{\boldsymbol{\sigma}}_{f,s}^\text{D}(n_{TX})\right)
    \label{eq:llrs_detection}
    \vspace{4pt} 
\end{equation}

\subsection{Training \gls{mdx}}

The proposed \gls{mdx} architecture is fully differentiable. It is trained by defining a loss function and back-propagating the gradients. We use \gls{bce} loss for the \glspl{llr} and \gls{mse} loss for the \gls{mimo} channel estimates. For a given estimate of \glspl{llr} $L$, and for every training data sample $n_{\text{TTI}}$ processed by \gls{mdx} (every TTI), the \gls{bce} loss function is defined as
\vspace{-6pt} 
\begin{align}
    \mathcal{J}^\text{BCE}_{n_{\text{TTI}}}(L) &= -\frac{1}{|\mathcal{D}| N_{TX} \prod_{n_{TX}=1}^{N_{TX}}B_{n_{TX}}} \sum_{n_{TX}=1}^{N_{TX}} \sum_{(f,s) \in \mathcal{D}} \sum_{b=1}^{B_{n_{TX}}} \notag \\ 
     &\Bigl(  q_{b,f,s,n_{TX}} \log_2 ( \hat{p}_{b,f,s,n_{TX}} ) + \notag \\ 
      &( 1-q_{b,f,s,n_{TX}} )  \log_2 ( 1-\hat{p}_{b,f,s,n_{TX}} ) \Bigr),
    \label{eq:loss_bce_detection}
    \vspace{-1pt} 
\end{align}
where $\hat{{p}}_{b,f,s,n_{TX}}=1/(1+\exp(-L_{b,f,s,n_{TX}}))$, $B_{n_{TX}}$ is the number of bits per symbol for the $n_{TX}$-th \gls{mimo} layer, and $|\mathcal{D}|$ represents the cardinality of the set $\mathcal{D}$. Similarly, a \gls{mse} loss is defined as
\vspace{-2pt} 
\begin{equation}
    \mathcal{J}^\text{MSE}_{n_{\text{TTI}}} (\hat{\mathbf{H}}_{f,s}) = \frac{1}{|\mathcal{D}||\mathcal{L}|} \sum_{(f,s) \in \mathcal{D}} \| \hat{\mathbf{H}}_{f,s} - \mathbf{H}_{f,s} \|_F^2,
    \label{eq:loss_mse}
\vspace{-6pt} 
\end{equation}
where $\hat{\mathbf{H}}_{f,s}$ is an estimate of the \gls{mimo} channel, and \(\|\cdot\|_F\) is the Frobenius norm.
In addition to using \gls{bce} loss on the final estimate of \glspl{llr} in \eqref{eq:llrs_detection}, we also put a \gls{bce} loss on an intermediate estimate of the \glspl{llr}. The intermediate estimate of \glspl{llr} are
\begin{equation}
    L_{b,f,s,n_{TX}}^{\text{DA-LS}}\! =\! \text{Demapper}_{1,\mathbf{1}} \left(\hat{\mathbf{x}}_{f,s}^\text{DA-LS}(n_{TX}), \hat{\boldsymbol{\sigma}}_{f,s}^\text{DA-LS}(n_{TX})\right),
    \label{eq:llrs_dals}
\end{equation}
where the subscripts "${1,\mathbf{1}}$" indicate that all learnable parameters of the demapping function are fixed to $1$, making them non-trainable. Then, the overall loss function can be defined over a training batch of size \( N_{\text{TTI}} \) as
\begin{align}
    \mathcal{J} =& \frac{1}{N_{\text{TTI}}} \sum_{n_{\text{TTI}}=1}^{N_{\text{TTI}}} \log_2( 1+\text{snr}_{n_{\text{TTI}}} ) \notag \\
    &\Bigl(   \mathcal{J}^\text{BCE}_{n_{\text{TTI}}}(L^\text{D}) + \mathcal{J}^\text{BCE}_{n_{\text{TTI}}}(L^\text{DA-LS}) + \lambda \cdot \mathcal{J}^\text{MSE}_{n_{\text{TTI}}} (\hat{\mathbf{H}}_{f,s}^{\text{NN}}) \Bigr),
    \label{eq_loss}
    \vspace{-5pt} 
\end{align}
where $\text{snr}_{n_{\text{TTI}}}$ is the linear SNR of the $n_{\text{TTI}}$-th training sample, and the multiplier $\lambda$ is a hyperparameter that weights the \gls{mse} loss. Note that the intermediate \glspl{llr} in \eqref{eq:llrs_dals} are computed only in the training phase.

\section{Simulation Results}

We compare the performance of \gls{mdx} with the \gls{soa} model from~\cite{cammerer2023neural,wiesmayr2024design}. In addition, three baseline methods are used for comparison: \gls{ls} channel estimation at pilot locations with linear interpolation to data symbols followed by \gls{lmmse} equalization, \gls{lmmse} channel estimation with K-best detection, and perfect channel knowledge with K-best detection. For more details on these baselines see~\cite{cammerer2023neural,wiesmayr2024design,hoydis2022sionna}.
\vspace{-15pt} 
\subsection{Simulation Setup}
\vspace{-1pt} 

We use the \gls{adam} optimization with a learning rate of $\text{lr}=0.001$, and $\lambda=0.01$ to train our \gls{mdx}. The end to end \gls{mimo} transmission is simulated in Sionna~\cite{hoydis2022sionna}. As in \cite{cammerer2023neural}, the training is done on the \gls{3gpp} \gls{umi} channel model, with random drops of users for each training data sample $n_{\text{TTI}}$, ensuring randomized power delay profiles, angle of arrival and departure. Also, the number of users, and their speeds are randomized: the number of active users is drawn from a triangular distribution, and the user speeds from a uniform distribution in $[0,56]$ m/s. The model was trained on 3 different modulation orders namely 4-QAM, 16-QAM, and 64-QAM with different \gls{ldpc} code rates corresponding to \gls{5g} \glspl{mcs} indices $I_{\text{MCS}}= \{9,14,19\}$ described in~\cite[Table 5.1.3.1-1]{3gpp_ts_38214}. 

All benchmarked methods employed an identical \gls{pusch} configuration: 4 \glspl{prb} for training and 273 for evaluation, with 2.14 GHz carrier frequency, 30 kHz subcarrier spacing. 
The DMRS configurations consist of 2 \gls{cdm} groups, with pilot symbols allocating \gls{ofdm} symbols 2, and 11. 
The evaluations are performed on \gls{mcs} indices $I_{\text{MCS}}$ and a \gls{tdl}-A channel model~\cite{3gpp_tr_38901}, with Doppler shifts and RMS delay spreads uniformly drawn from the intervals $[0,325]$ Hz and $[10,300]$ ns, respectively. 
The number of filters for \glspl{resblock} of \gls{mdx} is set to 8. All results in this paper were obtained using a single pre-trained \gls{mdx} model without additional training or fine-tuning. 
The \gls{mdx} model was trained on an NVIDIA GeForce GTX 1080 Ti GPU for 500,000 iterations with a batch size of $N_{\text{TTI}} = 128$.

\subsection{\gls{mu}-\gls{mimo} 4$\times$2}

\begin{figure*}[tbp!]
\centerline{\includegraphics[width=0.8\textwidth]{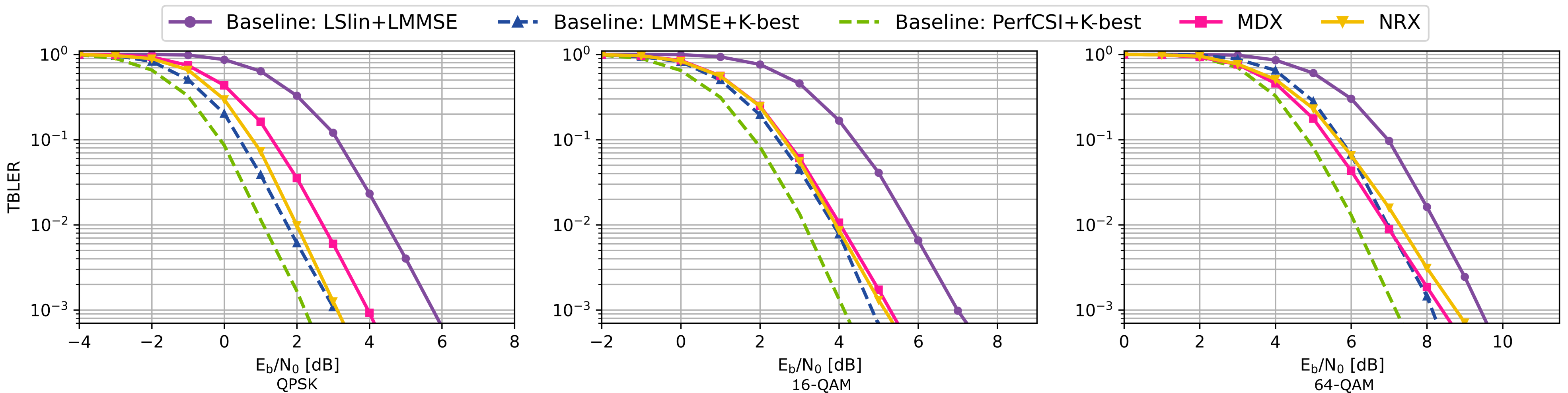}}
\vspace{-10pt} 
\caption{\gls{tbler} vs. \gls{snr} performance with $N_{TX}=2$ active layers, and $N_R=4$ receiving antenna elements for a 3GPP TDL-A channel. }
\vspace{-9pt} 
\label{fig:4x2}
\end{figure*}

We evaluated the receivers on a \gls{mu}-\gls{mimo} \gls{pusch} scenario with 2 active layers ($N_{TX}=2$), and 4 receiver antenna elements ($N_R=4$). 
Fig.~\ref{fig:4x2} shows the \gls{tbler} performance vs \gls{snr} values, for varying \gls{mcs} indices in $I_{\text{MCS}}$. 
For this evaluation we use the NRX model that supports varying \gls{mcs} by a masking method. We use the weights open sourced in~\cite{wiesmayr2024design}. 
Our model outperforms the \gls{ls} channel estimation baseline across all three modulation orders and achieves performance comparable to \gls{lmmse} channel estimation baseline at 16-QAM and 64-QAM. While \gls{mdx} surpasses NRX at 64-QAM, NRX performs better at QPSK, albeit at the cost of significantly higher computational complexity.

\subsection{\gls{mu}-\gls{mimo} 16$\times$4}

\begin{figure*}[tbp]
\centerline{\includegraphics[width=.9\textwidth]{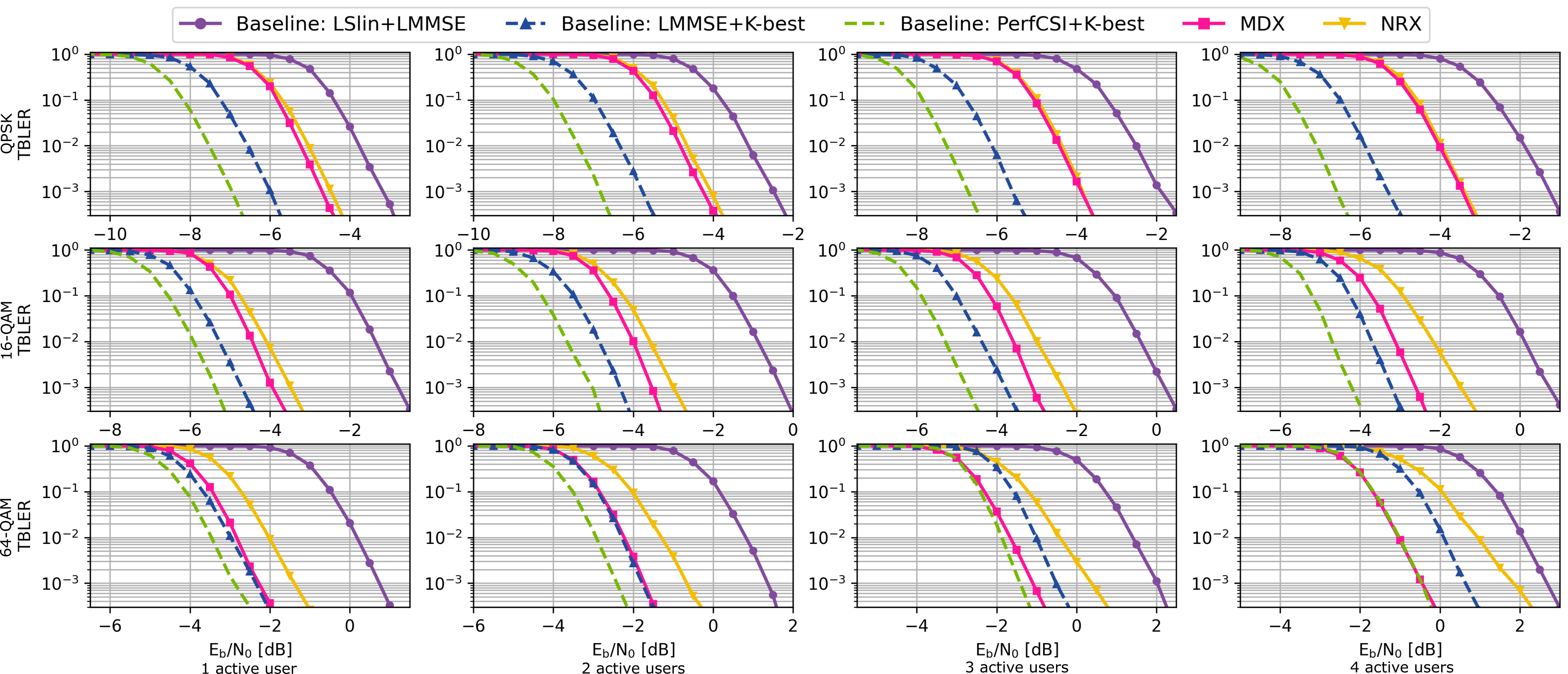}}
\vspace{-9pt} 
\caption{\gls{tbler} vs. SNR for varying \gls{mcs} indices and number of active layers in a 3GPP TDL-A channel. Columns represent the number of active layers, ranging from 1 (left) to 4 (right). Rows correspond to \gls{mcs} indices: 9 (top), 14 (middle), and 19 (bottom).}
\vspace{-16pt} 
\label{fig:16x4}
\end{figure*}

Fig.~\ref{fig:16x4} shows the evaluation results of the receivers on a \gls{mu}-\gls{mimo} \gls{pusch} scenario with a varying number of active layers (maximum $N_{TX}=4$), 16 receiver antenna elements ($N_R=16$), and for \gls{mcs} indices in $I_{\text{MCS}}$. 
Here we use the same \gls{mdx} model from previous evaluation (trained on \gls{mu}-\gls{mimo} 4$\times$2), without further training.

Using the source code from~\cite{cammerer2023neural,wiesmayr2024design}, we trained a separate NRX model for each of the three \gls{mcs} indices in $I_{\text{MCS}}$. We adopted the parameterization (e.g., filter counts, iterations) from \cite{cammerer2023neural}, and set the unspecified $d_s \! =\! 64$. Each model was trained on \gls{cscs} resources (NVIDIA GH200 GPUs) for several million iterations with a batch size of 128.

Our model outperforms the baseline with \gls{ls} channel estimation as well as NRX across all tested modulation orders and number of active layers. It performs close to the \gls{lmmse} channel estimation baseline at 16-QAM modulation, and approaches perfect channel knowledge at 64-QAM. 

\vspace{-3pt} 
\subsection{Complexity Analysis}
\vspace{-2pt} 

The complexity of \gls{mdx} is dominated by two \gls{lmmse} blocks, and \glspl{resblock} processing (mostly depth-wise separable convolutional neural networks). The NRX model uses compute-expensive depthwise separable convolutional layers, and fully connected layers. We count one complex-valued multiplication as four real-valued multiplications. The number of real valued multiplications for our \gls{lmmse} block is $2N_{TX}^3+6N_RN_{TX}^2+6N_RN_{TX}-2N_{TX}+2$ per \gls{re}. For the pair of a depthwise separable convolution followed by a point-wise convolution (with stride set to 1) we count $k^2N_0FS+N_0NFS$ multiplications ($k$, $N_0$, and $N$ denote the filter size, number of input feature maps, and number of output feature maps). The complexities, in \glspl{flop}, along with the number of parameters for the \gls{mdx} and NRX models, configured as for the \gls{tbler} evaluation, are in Table~\ref{tab:flops}. In the 4$\times$2 \gls{mu}-\gls{mimo} configuration, our \gls{mdx} model requires 106$\times$ fewer \glspl{flop} and 157$\times$ fewer parameters than the NRX model. For the 16$\times$4 \gls{mu}-\gls{mimo} configuration, these reductions are 66$\times$ and 396$\times$, respectively.

\begin{table}[ht]
\centering
\vspace{-9pt} 
\caption{FLOPs and Parameters}
\vspace{-5pt} 

\label{tab:flops}
\begin{tabular}{l@{\hspace{5pt}} l@{\hspace{5pt}} rrp{2cm}}
\hline

\textbf{MIMO} & \textbf{Model} & \textbf{FLOPs(G)} & \textbf{Params(k)} & \textbf{NRX/MDX} \\

\hline

$4\times2$ & \gls{mdx} & 0.7 & 2.7 & $\mathbf{106\times}$ (FLOPs) \\
 & NRX & 78.6 & 431.2 & $\mathbf{157\times}$ (Params) \\
\hline
$16\times4$ & \gls{mdx} & 6 & 2.7 & $\mathbf{66\times}$ (FLOPs) \\
 & NRX & 397.6 & 1,088.4 & $\mathbf{396\times}$ (Params) \\
\hline
\end{tabular}
\vspace{-6pt} 
\end{table}

\section{Conclusion}

This paper presents a \gls{5g} \gls{nr} \gls{pusch} receiver that integrates conventional \gls{phy} blocks with learnable parameters and \glspl{resblock} processing. By employing a model-driven approach, we substantially lowered the receiver’s computational complexity (by 66$\times$) and memory requirements (by 396$\times$), while surpassing the performance of both classical baselines and \gls{soa} solutions. 
Additionally, we simplified the training phase, which is essential for online learning using site-specific data captured from the actual radio environment at the edge base stations. On the other hand, our receiver processes each \gls{mu}-\gls{mimo} link independently using shared parameters and depthwise separable convolutions, unlocking significant potential for parallel processing. Furthermore, it demonstrates greater flexibility than existing \gls{soa} solutions, supporting a wide range of \gls{pusch} \gls{mu}-\gls{mimo} configurations—including variations in \gls{mimo} layers, receiver antennas, bandwidths, and modulation orders—using a single pre-trained model. 
The \gls{mdx} implementation along with trained weights will be open-sourced to ensure reproducibility~\cite{open_source}.
\vspace{-2pt} 

\section*{Acknowledgment}

This work was supported by a grant from the Swiss National Supercomputing Centre (CSCS) under project ID lp12 on Alps.

\vspace{-8pt} 

\Urlmuskip=0mu plus 1mu\relax
\def\UrlBreaks{\do\/\do-}
\bibliographystyle{IEEEtran}
\bibliography{IEEEabrv,bib}

\end{document}